\documentclass[useAMS,usenatbib,letters]{mnras}
\usepackage{amsmath}
\usepackage{graphicx}
\usepackage{xcolor}
\usepackage{hhline}
\usepackage{cancel}
\usepackage[normalem]{ulem}

\newcommand{\pcb}[1]{\textcolor{red}{{}{#1}}}

\newcommand{\be}{\begin{equation}}
\newcommand{\ee}{\end{equation}}
\newcommand{\Pn}{\mathcal{P}(N)}
\newcommand{\e}{\times10^}
\newcommand{\Mmin}{M_{\rm{min}}}
\newcommand{\TCO}{\left<T_{\rm{CO}}\right>}
\newcommand{\pa}{\partial}

\begin{document}

\title[SFR from intensity maps]{The high-redshift star formation history from carbon monoxide intensity maps}
\author[Patrick C. Breysse, Ely D. Kovetz, and Marc Kamionkowski]{Patrick C. Breysse,$^{1}$\thanks{pbreysse@pha.jhu.edu (PCB)} Ely D. Kovetz,$^{1}$\footnotemark[1] and Marc Kamionkowski$^{1}$\footnotemark[1] \\
$^{1}$ Department of Physics and Astronomy, Johns Hopkins University, Baltimore, MD 21218 USA}

\maketitle
\label{firstpage}

\begin{abstract}
We demonstrate how cosmic star-formation history can be measured with one-point statistics of carbon-monoxide intensity maps.  Using a P(D) analysis, the luminosity function of CO-emitting sources can be inferred from the measured one-point intensity PDF.  The star-formation rate density (SFRD) can then be obtained, at several redshifts, from the CO luminosity density.  We study the effects of instrumental noise, line foregrounds, and target redshift, and obtain constraints on the CO luminosity density of order 10\%.  We show that the SFRD uncertainty is dominated by that of the model connecting CO luminosity and star formation.  For pessimistic estimates of this model uncertainty, we obtain an error of order 50\% on SFRD for surveys targeting redshifts between 2 and 7 with reasonable noise and foregrounds included.  However, comparisons between intensity maps and galaxies could substantially reduce this model uncertainty.  In this case our constraints on SFRD at these redshifts improve to roughly $5-10\%$, which is highly competitive with current measurements.
\end{abstract}


\section{Introduction}
In recent years, our understanding of the high-redshift universe has dramatically improved thanks to a variety of multi-wavelength galaxy surveys. In particular, we have a much improved understanding of the history of star formation even in distant galaxies, indicating that the cosmic star formation rate density (SFRD) increases up to a maximum around $z\sim2-3$ before declining until the present day (\citet{madau} and references therein).  However, most of our knowledge of very high redshifts comes from a relatively small population of the brightest galaxies.  

In this Letter we will discuss intensity mapping, a new means to study the distribution and properties of high-redshift galaxies.  We will show how it is possible to obtain strong constraints on SFRD using this intensity mapping technique, in principle as strong as $\sim10\%$ for near future surveys targeting $z=2.6$.  This constraint does not diminish significantly at higher redshifts, which suggests that intensity mapping may be a powerful tool for constraining the high-redshift SFRD.

Instead of observing the emission of single galaxies, \emph{intensity mapping} involves measuring how the intensity of a single spectral line varies on relatively large scales.  With a beam much larger than a typical galaxy, an intensity mapping experiment measures the aggregate emission from a large number of galaxies, requiring neither the high sensitivity or high resolution of a traditional galaxy survey.  

Intensity mapping is commonly discussed using the 21 cm neutral hydrogen line, specifically targeted at the epoch of reionization \citep{mmr}. However in recent years observers have contemplated using other lines.  Proposed lines include CII \citep{gcs,S14} and Ly$\alpha$ \citep{pullenb,gong}, but we will consider here the 115 GHz CO(1-0) rotational transition.  CO is a commonly-observed molecule in nearby galaxies as it provides an excellent tracer of molecular gas, which in turn traces star formation activity \citep{Bolatto}.  In an intensity mapping experiment, CO could be used to measure star formation in high-redshift galaxies, without being restricted to the brightest sources.  At least one such experiment, the 
Carbon Monoxide Mapping Array Pathfinder, is currently in the planning stages \citep{li}.

The typical statistic used to study an intensity map is the power spectrum $P(k)$, which can be written as
\be
P_{\rm{CO}}(k,z) = \TCO^2(z)P_{\rm{gal}}(k,z)+P_{\rm{shot}}(z),
\ee
where $\TCO$ is the sky-averaged brightness temperature, $P_{\rm{gal}}$ is the galaxy power spectrum, and $P_{\rm{shot}}$ is scale-independent shot noise due to the discreteness of the CO sources.  Both $\TCO$ and $P_{\rm{shot}}$ depend sensitively on the properties of the galaxy population, but both depend only on integrals over the CO luminosity function, which limits their ability to constrain its shape.  On top of that, $\TCO$ is degenerate with the galaxy bias and the matter growth function, and $P_{\rm{shot}}$ may be degenerate with shot noise contributions from foreground lines such as HCN(1-0) \citep{FG}.

As the CO luminosity and star formation rate (SFR) of a galaxy may be related nonlinearly, these integral constraints are suboptimal for constraining SFRD.  What we would like is a full measurement of the CO luminosity function, which would allow us to make much better SFR estimates.  For this, we turn to the one-point statistics of the map.  Specifically, we use a technique known as P(D) analysis, first proposed by \citet{PofD}.  P(D) analysis is a method for obtaining source number counts in a confusion-limited survey.  Some recent examples of its use include studies of the cosmic infrared background \citep{hermes} and gamma-ray emission from our Galaxy \citep{PD,PD2}.  Below we will demonstrate how this technique can relate the one-point PDF of an intensity map to the underlying luminosity function and illustrate its potential for constraining high-redshift star formation.

\vspace{-0.15in}

\section{P(D) Analysis}

The one-point PDF of a CO intensity map depends both on the CO luminosity function and the clustering properties of the galaxy population.  These quantities are related by
\be
P(T)=\sum_{N=0}^\infty \Pn P_N(T),
\ee
where $P(T)$ is the probability of observing a pixel with brightness temperature $T$, $P_N(T)$ is the probability of observing intensity $T$ in a pixel containing $N$ sources, and $\Pn$ is the probability that a given pixel contains $N$ sources \citep{PD}.  The PDF for a pixel containing zero sources is a delta function, and the PDF for a pixel with a single source can be determined from the luminosity function.  For higher values of $N$, $P_N(T)$ can be determined through a series of convolutions:
\be
P_N(T)=\int_0^\infty P_{N-1}(T-T_1)P_1(T_1)dT_1.
\label{conv}
\ee

We then need a model for the luminosity function.  We follow \citet{FG} and assume that the CO luminosity $L$ of a halo is related to its mass $M$ by a power law, $L=AM^b$, where $A_{\rm{CO}}=2\e{-6}$ and $b_{\rm{CO}}=1$, and luminosity and mass are in Solar units.  We consider only halos more massive than $\Mmin=10^9$ Solar masses, and assume that only a fraction $f_{\rm{duty}}=t_{\rm{age}}/(10^8\ \rm{yr})$ of them are emitting CO at any given time, where $t_{\rm{age}}$ is the age of the universe in years.  The number density of halos with a given mass is then determined by a mass function $dn/dM$ \citep{tinker}.  If $\bar{n}$ is the mean number density of halos with masses greater than $\Mmin$ calculated by integrating this mass function, then we can write $P_1(M)=(1/\bar{n})dn/dM$.  We can then convert this to $P_1(L)$ using our mass-luminosity relation, and from there to a brightness temperature \citep{lidz}.

If we assume that the sources are unclustered, then the probability $\Pn$ is a Poisson distribution with mean $\mu=\bar{n}V_{\rm{pix}}$ for pixels with volume $V_{\rm{pix}}$.  However, the source clustering will alter the shape of this distribution \citep{barcons}.  A perfect understanding of the clustering would require knowledge of the full multi-point statistics of the source population. However we can obtain a reasonable estimate of the clustering by assuming the sources are distributed log-normally \citep{cj}.  For this work, we follow the assumptions made in \citet{FG} and assume that the number of sources $N$ in a pixel is determined by drawing from a Poisson distribution $P_{\rm{Poiss}}(N,\mu')$ with mean $\mu'$ which is in turn drawn from a log-normal distribution $P_{\rm{LN}}(\mu')$.  We then have
\be
\Pn=\int_0^\infty P_{\rm{LN}}(\mu')P_{\rm{Poiss}}(N,\mu')d\mu',
\ee
where
\be
P_{\rm{LN}}(\mu)=\frac{1}{\mu\sqrt{2\pi\sigma_G^2}}e^{-\left[\ln\left(\mu/\bar{n}\right)+\sigma_G^2/2\right]^2/2\sigma_G^2},
\ee
and $\sigma_G^2$ is the variance parameter from Equation (3.6) of \citet{FG}, which depends on the chosen form of the power spectrum (however it is of order unity and has a small impact on the final form of  P(T)).

It should be noted that the proof-of-concept model considered here is simplified in several respects for ease of computation.  For example, the duty cycle $f_{\rm{duty}}$ we use here is independent of halo mass, and we have a hard cutoff on SFR below $\Mmin$.  A more physical treatment would be to use a duty cycle that varies smoothly with mass and goes to zero at low masses (see for example \citet{fduty}).  In addition, a realistic CO luminosity function likely has a ``knee" at a fainter luminosity than our model predicts \citep{Obr}, meaning that our calculation may over-predict the number counts of the brightest sources.  It may be preferable to use a luminosity function model with more than two parameters which can accurately capture this ``knee", such as a Schechter function.  We leave exploration of these issues and their effects on our results to future work (Breysse et al., in preparation).

With Equations (1-3) we can calculate the one-point PDF of a CO intensity map.  However, a real map will include contributions from instrumental noise as well as foreground emission.  For this Letter we ignore the effects of continuum foregrounds such as dust and synchrotron emission, but we do consider foregrounds with line spectra \citep{FG}, as well as instrumental noise.  Specifically, we consider contamination from the 88 GHz HCN(1-0) line, which we model in the same way as CO, with $A_{HCN}=1.7\e{-15}$ and $b_{HCN} = 5/3$ (estimated based on \citet{HCN}).  We assume the instrumental noise has a Gaussian PDF with zero mean and variance $\sigma_N$.  These contaminants can be added to the PDF by convolving $P(T)$ for the original map with that of the contaminant.  

We base our model survey on the ``Full" CO mapping experiment described in Table 2 of \citet{li}.  This experiment surveys 6.25 square degrees between $z=2.4$ and 2.8, with $\sigma_N=5.8\ \mu$K.  For simplicity, we assume that neither the signal nor the foregrounds evolve significantly across the observed frequency range.  We use information from all of the frequency channels to study the CO properties averaged over the full survey volume.  

\begin{figure}
\centering
\includegraphics[width=\columnwidth]{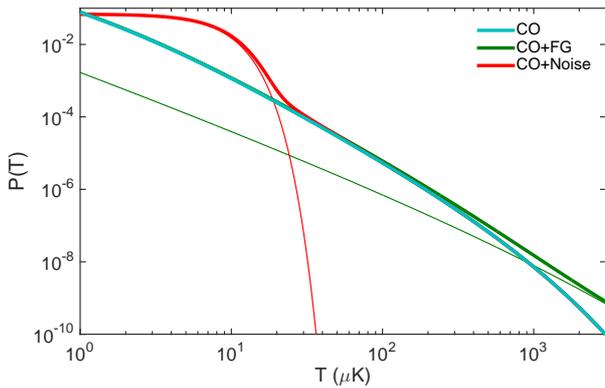}
\vspace{-0.3in}
\caption{Predicted PDFs for intensity maps of CO, CO with HCN, and CO with instrumental noise ({\it Thick curves}: the full PDF; {\it Thin curves}: the contaminant alone).}  
\vspace{-0.3in}
\label{hist}
\end{figure}

Figure \ref{hist} shows the predicted one-point PDFs for the CO signal and the two contaminants.  We use 50 logarithmically spaced bins between $T=2$ and 1000 $\mu$K, neglecting intensities outside of this range.  The number of pixels within each bin is computed by integrating $P(T)$ over the width of each bin.  Below $T\sim1\ \mu$K, unphysical ``ringing" effects come into the PDF due to the hard cutoff at $\Mmin$.  Since this regime would be noise dominated in a realistic experiment, the effect on the CO constraints should be small. 

\vspace{-0.15in}

\section{CO Luminosity Constraints}

In order to determine the constraining power of such an experiment, we perform a Fisher analysis on our calculated $P(T)$ curves.  We calculate the Fisher matrix
\be
F_{ij} = \sum_{k=1}^{N_{\rm{bin}}}\frac{1}{N_k}\frac{\pa N_k}{\pa X_i}\frac{\pa N_k}{\pa X_j},
\label{fisher}
\ee
where $N_k$ is the predicted number of pixels in bin $k$ and $X_i=(A_{\rm{CO}},\ b_{\rm{CO}},\ A_{HCN},\ b_{HCN})$ are the four parameters we fit for.  Note that we have assumed that the instrumental noise is well understood \emph{a priori}.  The derivatives in the Fisher calculation are evaluated at the maximum fiducial parameter values given above.  We have assumed Poisson uncertainties on the number of pixels in each bin, so the error on $N_k$ is $\sqrt{N_k}$.

\begin{figure}
\centering
\includegraphics[width=\columnwidth]{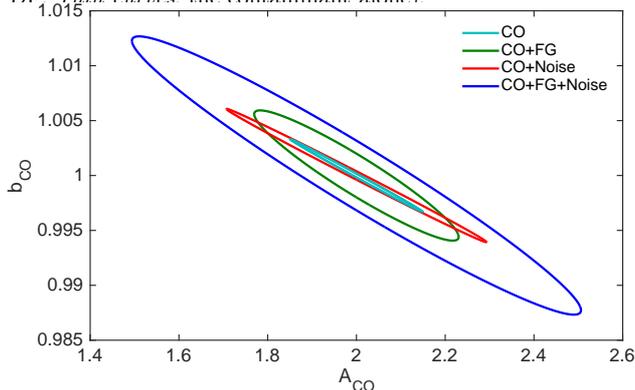}
\vspace{-0.3in}
\caption{Predicted 95\% confidence regions for CO mass-luminosity parameters $A_{\rm{CO}}$ and $b_{\rm{CO}}$ for four cases: only CO with no noise or foregrounds (light blue), CO with foreground HCN (green), CO with instrumental noise (red) and CO with both noise and foreground (blue).}
\vspace{-0.1in}
\label{Ab}
\end{figure}

We can invert the Fisher matrix to obtain the covariance matrix, and marginalize over the foreground parameters to plot confidence regions for our CO parameters.  Figure \ref{Ab} shows the 95\% confidence ellipses for a map containing just CO, as well as for maps contaminated by HCN and instrumental noise.   With no foreground lines, the parameters $A_{\rm{CO}}$ and $b_{\rm{CO}}$ are fairly degenerate, since to linear order increasing one or the other simply makes every halo brighter.  Even with both foregrounds and noise included, the uncertainties are around the $\sim20\%$ level in $A_{\rm{CO}}$ and the $\sim 1\%$ level in $b_{\rm{CO}}$, so relying on one-point statistics, these intensity maps can provide excellent constraints on the CO luminosity function.  For comparison, \citet{BKK} estimated an uncertainty on $A_{\rm{CO}}$ of roughly an order of magnitude.

Now that we have constraints on $A$ and $b$, we can calculate constraints on other astrophysically interesting quantities, such as $\TCO$, which in addition to being a useful astrophysical quantity in its own right, is important for understanding the power spectrum of an intensity map.  
The mean volume emissivity of CO emitters is simply
\be
j_{\rm{CO}}=A_{\rm{CO}}\int_{\Mmin}^\infty M^{b_{\rm{CO}}}\frac{dn}{dM}dM,
\ee
which can then be easily converted to brightness temperature \citep{lidz}.  

\vspace{-0.15in}

\section{Star Formation Constraints}
The focus of this Letter is on high-redshift star formation, so we will now demonstrate how the P(D) analysis described above allows us to constrain SFRD from the CO luminosity function.  Our mass-luminosity parameters $A$ and $b$ in \citet{FG} were originally derived from a set of empirical scaling relations discussed in \citet{pullen} and \citet{lidz}.  The CO and FIR luminosities of a galaxy have a well-known correlation which we write as
\be
\frac{L_{\rm{FIR}}}{L_{\sun}}=C_{\rm{FIR}}\left(\frac{L'_{\rm{CO}}}{\rm{K\ km\ s^{-1}\ pc^2}}\right)^{X_{\rm{FIR}}},
\label{fir}
\ee
where $C_{\rm{FIR}}$ and $X_{\rm{FIR}}$ are constants and $L_{\rm{CO}}/L_{\sun}=4.6\e{-5}(L'_{\rm{CO}}/\rm{K\ km\ s^{-1}\ pc^2})$ \citep{wang,cw}.  The FIR luminosity of a galaxy can then be related to its SFR through the Kenicutt relation
\be
\frac{\rm{SFR}}{M_{\sun}/\rm{yr}}=C_{\rm{SFR}}\frac{L_{\rm{FIR}}}{L_{\sun}},
\label{sfr}
\ee
for some constant $C_{\rm{SFR}}$ \citep{kennicutt}.  The values for the above constants\footnote{In the notation of \citet{li}, $\alpha = X_{\rm{FIR}}$, $\beta=\log(C_{\rm{FIR}})$, and $\delta_{\rm{MF}}=10^{10}C_{\rm{SFR}}$} used in \citet{FG} are $C_{\rm{FIR}}=1.35\e{-5}$, $X_{\rm{FIR}}=5/3$, and $C_{\rm{SFR}}=1.5\e{-10}$.  \citet{pullen} stated that the weakest part of this process is the relation between SFR and halo mass; here we assume they are related by a power law and use the above relations to write it in terms of $A$ and $b$:
\be
\rm{SFR}(M) = 9.8\e{-18}\left(\frac{A_{\rm{CO}}}{2\e{-6}}\right)M^{5b_{\rm{CO}}/3}.
\ee
We can then integrate this over the mass function to obtain the mean SFRD $\psi$ in our survey, which turns out to be 0.12 $M_{\sun}/\rm{yr}/\rm{Mpc}^3$ for our fiducial model.

\begin{figure}
\centering
\includegraphics[width=\columnwidth]{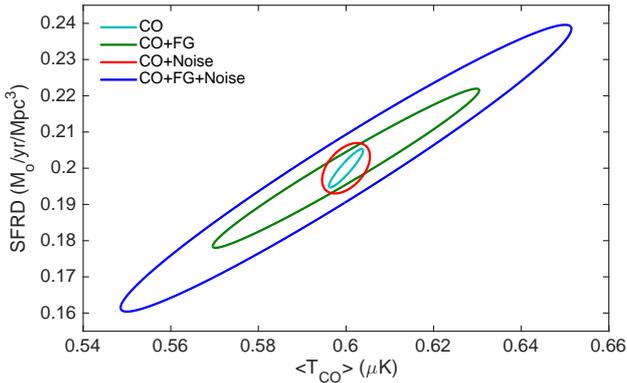}
\vspace{-0.3in}
\caption{Predicted 95\% confidence regions for mean CO brightness temperature $\TCO$ and SFRD for the same four cases shown in Figure \ref{Ab}.}
\vspace{-0.3in}
\label{ST}
\end{figure}

Constraints on our original parameters $X_i$ are straightforward to convert to constraints on other parameters $Y_i$.  We simply need to multiply the Fisher matrix calculated in Equation (\ref{fisher}) on both sides by the Jacobian matrix $J_{ij} = \partial X_i/\pa Y_j$.  If $Y_i=(\TCO,\psi)$, we get the confidence regions plotted in Figure \ref{ST}.  The uncertainty on SFRD with noise and foregrounds included is on the order of $\sim10\%$.  It should be noted that this is an optimistic calculation.  When converting our Fisher matrix to SFRD and $\TCO$, we assumed that the scaling relations used to connect $L_{\rm{CO}}$ to SFR are well constrained.  In practice, the uncertainties in these relations will reduce the constraining power of such a measurement.  More galaxy observations will be required to reach the constraints shown here.  However, it is clear from these calculations that the potential of a CO intensity mapping experiment to constrain SFRD is quite high.

Up until now we have been considering a survey centered at $z=2.6$.  It is interesting to consider how the SFRD constraint varies with redshift, especially as the reliability of other methods diminishes with redshift.  Figures 4 and 5 illustrate this redshift dependence.  For this calculation, we hold all of the instrumental parameters constant except the central observing frequency.  Figure 4 shows the fractional uncertainty in SFRD as a function of redshift.  Figure 5 places the result with noise and foregrounds in context with results from the far ultraviolet (FUV) observations described in \citet{madau} and gamma-ray burst (GRB) observations described in \citet{beacom}.  For Figure 5, we use a somewhat different form for $\psi(z)$ than we have used thus far.  Our modeling, which is quite simplistic and only intended to serve as a proof-of-concept, gives a form for $\psi(z)$ (dotted black line) which differs from the \citet{madau} fit (solid black line).  To facilitate comparison with the data (grey points), the blue curves in Figure 5 show the 1-$\sigma$ uncertainty on SFRD calculated assuming the \citet{madau} form for $\psi(z)$.  

As mentioned above, there is theoretical uncertainty regarding the relations in Equations (\ref{fir}) and (\ref{sfr}).  The magenta curves in Figures 4 and 5 show the effect of taking it into account.  The dashed magenta curves in both figures assume a 10\% uncertainty on $C_{\rm{FIR}}$ and $C_{\rm{SFR}}$.  Current results such as \citet{cw} tend to give a constraint on $X_{\rm{FIR}}$ which is roughly an order of magnitude better than the constraint on $C_{\rm{FIR}}$, similar to how in Figure 2 we obtain a much better fractional constraint on $b_{\rm{CO}}$ than $A_{\rm{CO}}$.  These curves therefore assume a 1\% uncertainty on $X_{\rm{FIR}}$.  These constraints are somewhat better than current values, but they are likely pessimistic compared to what will be available once intensity mapping data are available.  If there are resolved CO emitters in the same volume, one could calibrate the CO-SFR relation and dramatically reduce this uncertainty.  The solid magenta curve in Figure 4 shows a more optimistic scenario where the errors on $C_{\rm{FIR}}$ and $C_{\rm{SFR}}$ are 1\% and the that on $X_{\rm{FIR}}$ is 0.1\%.

In order to test the model dependence of our results, we consider two aspects of more sophisticated CO emission models which have been neglected thus far in our deliberately simplified analysis.  First, results such as those of \citet{Obr} show that the CO luminosity function cuts off considerably earlier than the halo mass function.  In \citet{li}, this effect comes into play because the $L_{\rm{CO}}(M)$ relation turns over at a certain mass.  We take this into account by adding an exponential cutoff to our $L(M)$ at \pcb{a halo mass of} $M_*=2\e{12}$ solar masses, approximately where the \citet{li} turnover appears.  Secondly, semianalytic models such the one presented in \citet{Lagos} and \citet{popping} show that the CO luminosity of a galaxy depends on many parameters besides its mass.  Again following \citet{li}, we add lognormal scatter with $\sigma=0.3$ dex to our $L(M)$ model to account for this, preserving some dependence on mass while allowing for fluctuations of other galaxy properties.

The dotted curves in Figure 4 show the effects of these changes for the cases with and without instrumental noise.  We can see that increasing the model complexity and suppressing the bright end of the PDF has relatively little effect at lower redshifts but noticeably worsens the constraints on SFR at high redshift, up to roughly a factor of 5 worse at $z=7$.  However, even at these high redshifts the overall uncertainty would still be dominated by the error on the CO-SFR conversion seen in the magenta curves.

\begin{figure}
\centering
\includegraphics[width=\columnwidth]{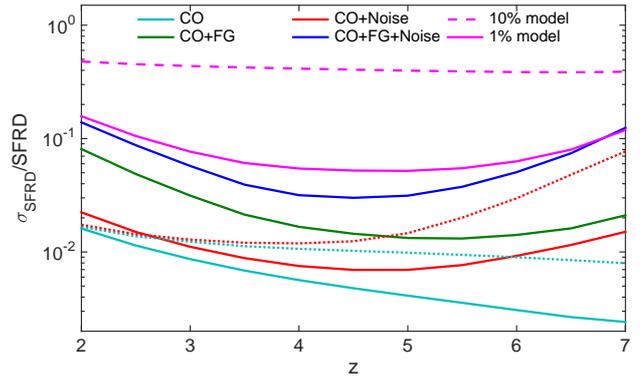}
\vspace{-0.2in}
\caption{Predicted fractional 1-$\sigma$ uncertainties on SFRD for different target redshifts.  The cyan, green, red, and blue curves show the same four cases as Figures 2 and 3.  The dashed magenta curve shows the effect of adding a pessimistic 10\% uncertainty on $C_{\rm{FIR}}$ and $C_{\rm{SFR}}$ and 1\% uncertainty on $X_{\rm{FIR}}$.  The solid magenta curve shows the effect of reducing these model uncertainties to a more optimistic 1\% and 0.1\% respectively.  The dotted cyan and red curves show the effect of adding an exponential cutoff and scatter to the power law $L(M)$ model as discussed at the end of Section 4.  These changes have little effect at low redshift, and reduce the constraint by a factor of $\sim5$ at high redshift.}
\vspace{-0.15in}
\end{figure}

\begin{figure}
\centering
\includegraphics[width=\columnwidth]{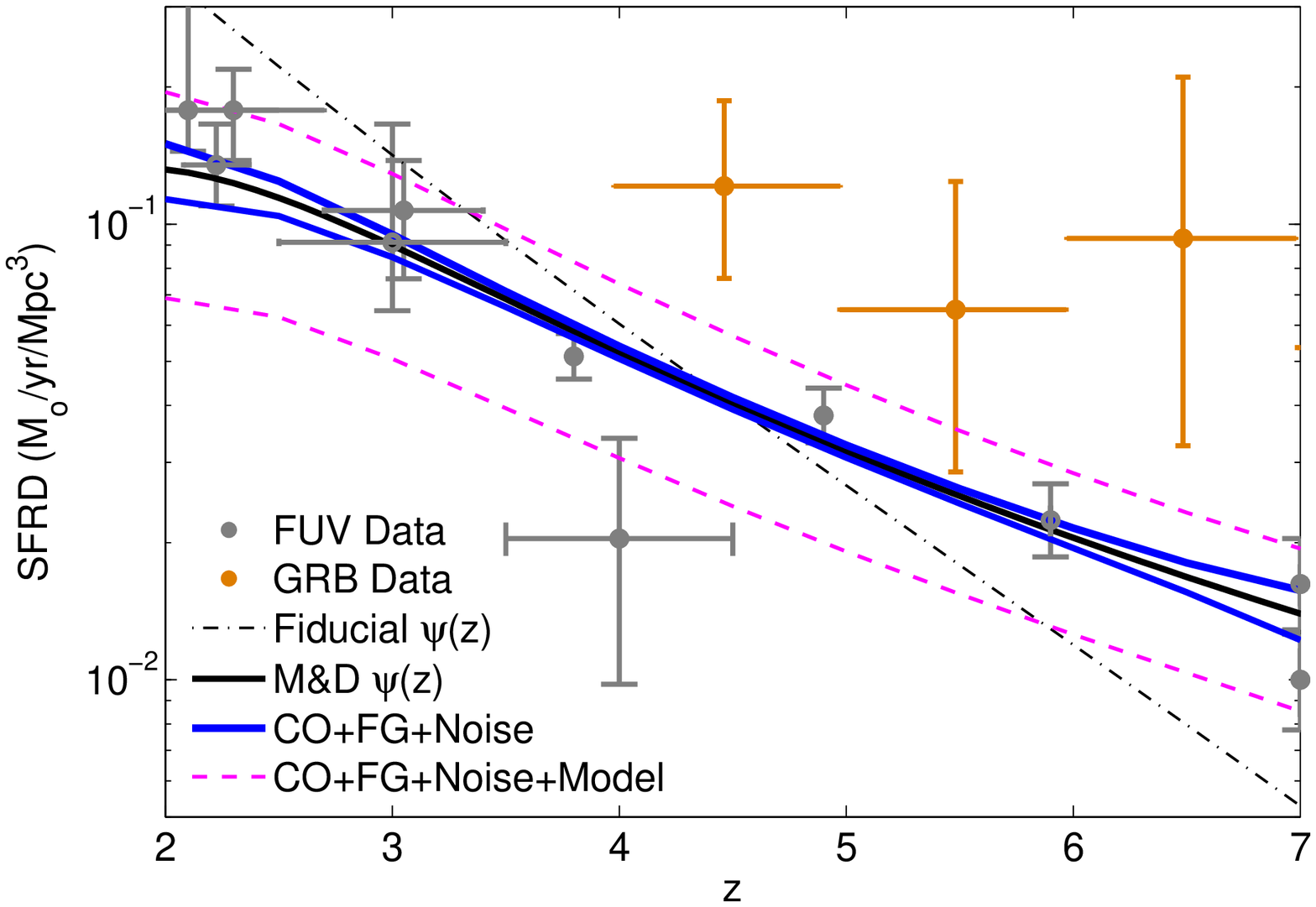}
\vspace{-0.3in}
\caption{Comparison between our predicted SFRD constraints and existing FUV data from \citet{madau} (grey points) and GRB data from \citet{beacom} (orange points).  The dash-dotted black line shows our simplistic fiducial $\psi(z)$ model, and the solid black curve shows the \citet{madau} fit to the data.  The blue curves show the $\pm1\sigma$ uncertainty from CO intensity mapping calculated assuming the fitted $\psi(z)$, including foregrounds and noise, but excluding modeling uncertainty.  The dashed magenta curves show the effect of adding the pessimistic 10\% model uncertainty from the dashed magenta curve in Figure 4.}
\vspace{-0.25in}
\end{figure}

\vspace{-0.2in}

\section{Conclusions}

From the above results we can get a good sense of the potential of this technique for constraining SFRD.  In an ideal world with no foregrounds, noise, or modeling uncertainty, this 6.25 deg$^2$ survey could constrain SFRD to $\sim1\%$ at $z\sim3$.  This scenario is obviously unrealistic, but we could reduce the modeling uncertainty with more observations and reduce the noise with more sensitive instruments.  It may be possible to reduce the foreground contamination as well.  The masking procedure described in \citet{FG} will have little effect on the one-point statistics since it would simply involve ignoring all pixels greater than some $T_{\rm{max}}$.  However, as described for Ly$\alpha$ intensity mapping in \citet{gong}, it may be possible to reduce the foreground levels by using other data from the same area of the sky.

Counterintuitively, the constraint on SFRD in Figure 4 seems to improve with redshift when foregrounds and noise are neglected.  This is because at higher redshifts, the same survey parameters correspond to a larger volume of space, which will include more sources.  In our simple model, $A_{\rm{CO}}$ and $b_{\rm{CO}}$ do not evolve with redshift, so the only effect of moving to more distant sources comes from the evolution of the mass function.  Eventually there will be few enough sources above $\Mmin$ that the constraint will worsen, but for our redshift range the volume increase outweighs this effect.  When noise is included, the volume effect dominates at low redshift, but the amplitude of the noise relative to $\TCO$ takes over at $z\sim5$ or so.  With foregrounds, the relative amplitude of CO and HCN decreases with redshift, starting to dominate over volume effects at around $z\sim6$.  The model uncertainties as we have considered them here are nearly redshift independent, so the constraint with these included depends weakly on redshift.

As seen in the dotted curves in Figure 4, our constraints do not change drastically when we consider a more sophisticated model.  The model used for the dotted curves contains significant scatter around the mean $L(M)$ relation and adds a sharp cutoff to the luminosity of high mass halos, yet the change in fractional uncertainty on SFR is only order unity.  Though even this more intricate model is still simpler than that of \citet{li} or state of the art semianalytic models, our qualitative conclusions should still hold.  We leave a detailed quantitative prediction for future work.

Currently there is substantial disagreement between high redshift SFRD's measured using different methods, as illustrated by the roughly order-of-magnitude disagreement between the FUV and GRB data plotted in Figure 5.  \citep{beacom} found that this discrepancy can be explained if the FUV analysis underpredicts the quantity of low-luminosity galaxies.  Since intensity mapping is far more sensitive to the fainter population than other methods it could provide a powerful means to resolve this discrepancy.

An effect we have not taken into account here is contamination from continuum foregrounds. Though in principle these should be easy to clean out, it is unclear what form the residual contamination would take and how it would alter our results.   Another effect worth considering is the dependence of these results on the beam size.  It seems plausible that it would be easier to constrain the shape of the CO luminosity function with a smaller beam, since there would be fewer pixels with multiple galaxies.  We leave consideration of both of these effects for future work.

It may also be possible to constrain other quantities related to star formation with these intensity maps besides global SFRD.  For example, because the convolutions in Equation (\ref{conv}) depend on the full range of possible luminosities, the one-point PDF at all intensities depends on the chosen value of $\Mmin$.  Thus, even though the faintest pixels will be noise-dominated, it could be possible to determine $\Mmin$ from our PDF.  Another common unknown for high-redshift galaxies is what is the relative contribution to the overall intensity of high-mass vs. low-mass galaxies.  This question is hard to answer with galaxy surveys, but it should be more tractable with intensity mapping.  The methods described in this Letter should be readily applicable to these and many other problems of high-redshift star formation.

From Figure 5, it is clear that CO intensity mapping has great potential for constraining the high-redshift SFRD.  Though we are currently limited by modeling uncertainty, this should improve with better models and more observations. The potential bound shown by the blue curves, though idealized, is remarkably strong despite using only a small survey area.  In addition, the results shown above do not include any additional constraints from the two-point statistics of a map.  Though they neglect all possible line foregrounds, \citet{li} show that strong constraints on the CO luminosity function can be obtained from the power spectrum alone.  Combining intensity mapping with existing multiwavelength galaxy surveys should further improve the constraints.  More work is necessary before CO intensity mapping can be used to accurately determine SFRs, but these results clearly demonstrate the value of this effort.

MK acknowledges the hospitality of the Aspen Center for Physics, supported by NSF Grant No.\ 1066293.  The authors thank John Beacom, Peter Behroozi and Graeme Addison for useful discussions.  This work was supported by the John Templeton Foundation, the Simons Foundation, NSF grant PHY-1214000, and NASA ATP grant NNX15AB18G.


\vspace{-0.2in}

\begin{thebibliography}{56}

\bibitem[\protect\citeauthoryear{Barcons}{1992}]{barcons} 
Barcons X., 1992, ApJ, 396, 460 

\bibitem[\protect\citeauthoryear{Bolatto et al.}{2013}]{Bolatto} Bolatto A.~D., Wolfire M., Leroy A.~K., 2013, ARA\&A, 51, 207

\bibitem[\protect\citeauthoryear{Breysse et al.}{2014}]{BKK} Breysse P. C., Kovetz E. D., Kamionkowski M., 2014, MNRAS, 443, 3506

\bibitem[\protect\citeauthoryear{Breysse et al.}{2015}]{FG} Breysse P.~C., Kovetz E.~D., Kamionkowski M., 2015, MNRAS, 452, 3408 

\bibitem[\protect\citeauthoryear{Carilli 
\& Walter}{2013}]{cw} Carilli C.~L., Walter F., 2013, ARA\&A, 51, 105 

\bibitem[\protect\citeauthoryear{Coles \& Jones}{1991}]{cj} Coles P., Jones B., 1991, MNRAS, 248, 1

\bibitem[\protect\citeauthoryear{Glenn et al.}{2010}]{hermes} 
Glenn J., et al., 2010, MNRAS, 409, 109 

\bibitem[\protect\citeauthoryear{Gao 
\& Solomon}{2004}]{HCN} Gao Y., Solomon P.~M., 2004, ApJ, 606, 271

\bibitem[\protect\citeauthoryear{Gong et al.}{2012}]{gcs} Gong Y., Cooray A., Silva M., Santos M.~G., Bock J., Bradford C.~M., Zemcov M., 2012, ApJ, 745, 49 

\bibitem[\protect\citeauthoryear{Gong et al.}{2014}]{gong} 
Gong Y., Silva M., Cooray A., Santos M.~G., 2014, ApJ, 785, 72

\bibitem[\protect\citeauthoryear{Jaacks et al.}{2012}]{fduty} Jaacks J., Nagamine K., Choi J.~H., 2012, MNRAS, 427, 403

\bibitem[\protect\citeauthoryear{Kennicutt}{1998}]{kennicutt} 
Kennicutt R.~C., Jr., 1998, ApJ, 498, 541 

\bibitem[\protect\citeauthoryear{Kistler et 
al.}{2009}]{beacom} Kistler M.~D., Y{\"u}ksel H., Beacom 
J.~F., Hopkins A.~M., Wyithe J.~S.~B., 2009, ApJ, 705, L104 

\bibitem[\protect\citeauthoryear{Lagos et al.}{2012}]{Lagos} Lagos C.~d.~P., Bayet E., Baugh C.~M., Lacey C.~G., Bell T.~A., Fanidakis N., Geach J.~E., 2012, MNRAS, 426, 2142 

\bibitem[\protect\citeauthoryear{Lee et al.}{2009}]{PD} Lee S.~K., Ando S., Kamionkowski M., 2009, JCAP, 7, 007

\bibitem[\protect\citeauthoryear{Lee et al.}{2015}]{PD2} Lee S.~K., Lisanti M., Safdi B.~R., 2015, JCAP, 5, 056 

\bibitem[\protect\citeauthoryear{Li et al.}{2015}]{li} Li T.~Y., Wechsler R.~H., Devaraj K., Church S.~E., 2015, arXiv:1503.08833 

\bibitem[\protect\citeauthoryear{Lidz et al.}{2011}]{lidz} Lidz A., Furlanetto S.~R., Oh S.~P., Aguirre J., Chang T.-C., Dor{\'e} O., Pritchard J.~R., 2011, ApJ, 741, 70 

\bibitem[\protect\citeauthoryear{Madau 
\& Dickinson}{2014}]{madau} Madau P., Dickinson M., 2014, ARA\&A, 52, 415

\bibitem[\protect\citeauthoryear{Madau et al.}{1997}]{mmr} Madau P., Meiksin A., Rees M.~J., 1997, ApJ, 475, 429

\bibitem[\protect\citeauthoryear{Obreschkow et 
al.}{2009}]{Obr} Obreschkow D., Heywood I., Kl{\"o}ckner 
H.-R., Rawlings S., 2009, ApJ, 702, 1321 

\bibitem[\protect\citeauthoryear{Popping et al.}{2014}]{popping} Popping G., P{\'e}rez-Beaupuits J.~P., Spaans M., Trager S.~C., Somerville R.~S., 2014, MNRAS, 444, 1301 

\bibitem[\protect\citeauthoryear{Pullen et al.}{2013}]{pullen}  Pullen A.~R., Chang T.-C., Dor{\'e} O., Lidz A., 2013, ApJ, 768, 15 

\bibitem[\protect\citeauthoryear{Pullen et al.}{2014}]{pullenb} Pullen A.~R., Dor{\'e} O., Bock J., 2014, ApJ, 786, 111 

\bibitem[\protect\citeauthoryear{Scheuer}{1957}]{PofD} 
Scheuer P.~A.~G., 1957, PCPS, 53, 764 

\bibitem[\protect\citeauthoryear{Silva et al.}{2014}]{S14} 
Silva M.~B., Santos M.~G., Cooray A., Gong Y., 2014, arXiv:1410.4808

\bibitem[\protect\citeauthoryear{Tinker et al.}{2008}]{tinker} 
Tinker J., Kravtsov A.~V., Klypin A., Abazajian K., Warren M., Yepes G., 
Gottl{\"o}ber S., Holz D.~E., 2008, ApJ, 688, 709 


\bibitem[\protect\citeauthoryear{Wang et al.}{2010}]{wang} 
Wang R., et al., 2010, ApJ, 714, 699 

\end{thebibliography}
\end{document}